\documentclass[conference]{IEEEtran}
\IEEEoverridecommandlockouts
\usepackage{pgfplots}
\usepackage{tikz}
\usetikzlibrary{calc}
\makeatletter
\newcommand{\gettikzxy}[3]{%
  \tikz@scan@one@point\pgfutil@firstofone#1\relax
  \edef#2{\the\pgf@x}%
  \edef#3{\the\pgf@y}%
}
\usetikzlibrary{spy,backgrounds}

\usepackage{acronym}
\acrodef{6g}[6G]{the sixth generation}
\acrodef{bs}[BS]{base station}
\acrodef{bse}[BSE]{beam squint effect}
\acrodef{cp}[CP]{cyclic prefix}
\acrodef{elaa}[ELAA]{extremely large antenna array}
\acrodef{ff}[FF]{far-field}
\acrodef{las}[L\&S]{localization and sensing}
\acrodef{los}[LOS]{line-of-sight}
\acrodef{nf}[NF]{near-field}
\acrodef{ris}[RIS]{reconfigurable intelligent surface}
\acrodef{rtt}[RTT]{round-trip-time}
\acrodef{sns}[SNS]{spatial non-stationarity}
\acrodef{swm}[SWM]{spherical wave model}

\acrodef{siso}[SISO]{single-input-single-output}
\acrodef{mimo}[MIMO]{multi-input-multi-output}
\acrodef{ue}[UE]{user equipment}
\acrodef{dmimo}[D-MIMO]{distributed MIMO}
\acrodef{sp}[SP]{scatter point}
\acrodef{awgn}[AWGN]{additive white Gaussian noise}
\acrodef{nlos}[NLOS]{non-line-of-sight}
\acrodef{ofdm}[OFDM]{orthogonal frequency division multiplexing}
\acrodef{tdoa}[TDOA]{time-difference-of-arrival}
\acrodef{toa}[TOA]{time-of-arrival}
\acrodef{am}[AM]{artificial multipath}
\acrodef{an}[AN]{artificial noise}
\acrodef{csi}[CSI]{channel state information}
\acrodef{mcrb}[MCRB]{misspecified Cramér-Rao bound}
\acrodef{crb}[CRB]{Cramér-Rao bound}
\acrodef{lb}[LB]{lower bound}
\acrodef{rmse}[RMSE]{root mean squared error}
\acrodef{psd}[PSD]{power spectral density}
\acrodef{pdf}[PDF]{probability distribution function}
\acrodef{aoa}[AOA]{angle-of-arrival}
\acrodef{aod}[AOD]{angle-of-departure}
\acrodef{fim}[FIM]{Fisher information matrix}
\acrodef{crb}[CRB]{Cramér-Rao bound}
\acrodef{moo}[MOO]{multi-objective optimization}
\acrodef{qos}[QoS]{quality of service}
\acrodef{sdp}[SDP]{semi-definite programming}
\acrodef{lmi}[LMI]{linear matrix inequality}
\acrodef{sdr}[SDR]{semi-definite relaxation}
\acrodef{rcs}[RCS]{radar cross section}
\acrodef{isac}[ISAC]{integrated sensing and communication}
\acrodef{bp}[BP]{bistatic positioning}
\acrodef{ms}[MS]{monostatic sensing}
\acrodef{slam}[SLAM]{simultaneous localization and mapping}
\acrodef{ms}[MS]{monostatic sensing}
\acrodef{fdb}[FDB]{full-dimensional beamforming}
\acrodef{cpa}[CPA]{codebook-based power allocation}
\acrodef{apa}[APA]{average power allocation}

\usepackage{booktabs} 
\usepackage{amsmath}
\usepackage{graphicx} 
\usepackage{epstopdf}
\usepackage{amssymb}
\usepackage{amsfonts}
\usepackage{amsthm}
\usepackage{cite}
\usepackage{bm,comment}
\usepackage{algorithm}
\usepackage{algpseudocode}

\usepackage{subeqnarray}
\usepackage{subfigure}
\usepackage{multicol}
\usepackage{multirow}
\usepackage{diagbox}
\usepackage{slashbox}
\usepackage{stfloats}
\usepackage{float}
\usepackage{color} 
\usepackage{cases}
\usepackage{lipsum}

\usepackage{geometry}
\allowdisplaybreaks
\begin{document}
\setlength{\textfloatsep}{4pt}

\bstctlcite{IEEEexample:BSTcontrol}
\title{Optimized Beamforming for Joint Bistatic Positioning and Monostatic Sensing}
\author{Yuchen Zhang\IEEEauthorrefmark{1}, 
Hui Chen\IEEEauthorrefmark{2}, 
Pinjun Zheng\IEEEauthorrefmark{3}, 
Boyu Ning\IEEEauthorrefmark{4}, 
Henk Wymeersch\IEEEauthorrefmark{2}, 
and Tareq Y. Al-Naffouri\IEEEauthorrefmark{1}

\\
\IEEEauthorrefmark{1}King Abdullah University of Science and Technology, KSA
\ \ \ 
\IEEEauthorrefmark{2}Chalmers University of Technology, Sweden
\\
\IEEEauthorrefmark{3}University of British Columbia, Canada
\ \ 
\IEEEauthorrefmark{4}University of Electronic Science and Technology of China, China
\\
(yuchen.zhang@kaust.edu.sa)


\thanks{This work was supported 
in part by the SNS JU project 6G-DISAC under the EU’s Horizon Europe research, by innovation programme under Grant Agreement No 101139130, and by the Swedish Research Council through the project HAILS under VR Grant 2022-03007, and by the King Abdullah University of Science and Technology (KAUST) Office of Sponsored Research (OSR) under Award ORA-CRG2021-4695.
}
}
\maketitle

\begin{abstract}
We investigate the performance tradeoff between \textit{bistatic positioning (BP)} and \textit{monostatic sensing (MS)} in a multi-input multi-output orthogonal frequency division multiplexing scenario. We derive the Cramér-Rao bounds (CRBs) for BP at the user equipment and MS at the base station. To balance these objectives, we propose a multi-objective optimization framework that optimizes beamformers using a weighted-sum CRB approach, ensuring the weak Pareto boundary. We also introduce two mismatch-minimizing approaches, targeting beamformer mismatch and variance matrix mismatch, and solve them distinctly. Numerical results demonstrate the performance tradeoff between BP and MS, revealing significant gains with the proposed methods and highlighting the advantages of minimizing the weighted-sum mismatch of variance matrices.
\end{abstract}

\IEEEpeerreviewmaketitle
\section{Introduction}
Thanks to its inherent dual functionality, \ac{isac} is expected to be a cornerstone in 6G development, enabling modern wireless networks to incorporate sensing capabilities\cite{fan2022jsac}. Sensing refers to the ability of a network to detect and interpret its surroundings, while positioning is a critical component that involves determining the location of objects or users. 
By sharing the same infrastructure, accurate radio positioning can not only support communications but also support various location-information-driven applications, particularly in scenarios with poor satellite visibility\cite{hui2022st}, which is anticipated to drive promising applications such as massive twinning, autonomous driving, immersive telepresence, and more.

\begin{figure}[t]
\centering
\includegraphics[width=0.75\linewidth]{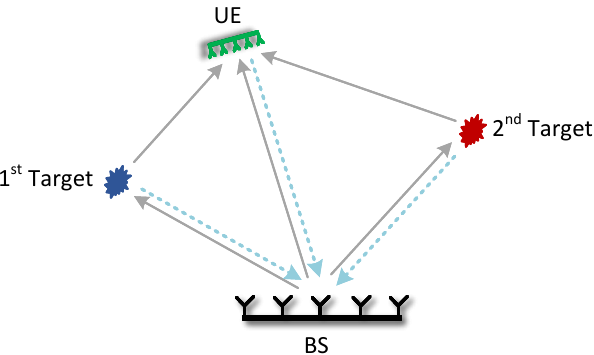}
\caption{Illustration of joint BP and MS, where the BS transmits pilot signals, functioning as a monostatic radar to sense passive targets and the UE. Meanwhile, the UE uses the received pilot signals to position itself.}\label{sys_mod}
\end{figure}

Radio positioning can be implemented in either bistatic or monostatic modes. In \ac{bp}, the transmitter and receiver are located at different positions, creating additional signal processing challenges, particularly due to the need for synchronization (or orientation estimation if the \ac{ue} has an antenna array). A common application of \ac{bp} in cellular networks involves \acp{ue} estimating their positions from pilots sent by \acp{bs}\cite{henk2019twc}. In contrast, radar-like \ac{ms} uses co-located transmitters and receivers, simplifying synchronization and signal processing, allowing cellular networks (e.g., \acp{bs}) to position environmental objects such as vehicles.

As ISAC evolves toward 6G, the increased frequency and path loss in communication systems necessitate high beamforming gain, which requires the use of large antenna arrays~\cite{hui2022st}. Beamforming optimization must not only consider communication performance but also account for positioning and sensing objectives. For example, traditional directional beamforming techniques are not optimal for positioning purposes~\cite{furkan2022tvt}. Given that ISAC systems share resources between sensing and communication, significant research has been dedicated to exploring the tradeoff between these two functions \cite{fan2018tsp,fan2018twc,fan2022tsp} and \ac{moo} problems can be formulated. Another way of dealing with the ISAC tradeoff is to maximize sensing performance while guaranteeing the minimum required
SINR of communication users~\cite{hua2023optimal}. In~\cite{he2023full}, the existing ISAC beamforming design is extended to a general case by considering the full-duplex capability. Usually, the beam design involves the formulation of a convex (or convexified) problem, which may not work well in reality with impairment, making learning-based approaches favorable~\cite{rivetti2023spatial}. 

In the above-mentioned works, bistatic positioning (BP) and monostatic sensing (MS) are usually discussed individually, and the complementary advantages of their co-existence are largely ignored. The authors in \cite{yu2023globecom} initiated the exploration by integrating \ac{bp} and \ac{ms} from a \ac{slam} perspective. However, no existing works have been identified to design beamformers on the tradeoff between these two paradigms.
This paper considers a \ac{mimo} \ac{ofdm} scenario and characterizes the performance tradeoff between \ac{bp} and \ac{ms} by designing beamforming judiciously. The key contributions are summarized as follows.
\begin{enumerate}
    \item We systematically derive the \acp{crb} for \ac{bp} and \ac{ms} as functions of beamformers, addressing distinct objectives: the \ac{ue} estimating its own position (considering clock bias and orientation mismatch) and the \ac{bs} sensing and positioning passive targets (including the \ac{ue}); 
    \item To strike a tradeoff between \ac{bp} and \ac{ms}, we formulate a \ac{moo} problem for beamforming design. A weighted-sum \ac{crb} approach is proposed to ensure a weak Pareto boundary. Additionally, weighted-sum mismatch-minimizing approaches, based on beamformer mismatch and variance matrix mismatch, are introduced and solved using distinct techniques; 
    \item We provide comprehensive numerical results that reveal the tradeoff between \ac{bp} and \ac{ms}. Specifically, compared to other baselines, the proposed approaches demonstrate significant performance gains. Furthermore, we highlight the superiority of minimizing the weighted-sum mismatch of variance matrices over the beamformer.
\end{enumerate}


\section{System Model And Problem Formulation} 
\subsection{Signal Model}
As illustrated in Fig. \ref{sys_mod}, we consider a \ac{mimo} \ac{ofdm}-based joint \ac{bp} and \ac{ms} system with $M$ subcarriers, where a \ac{bs} equipped with $M_{\text{T}}$ transmit antennas transmits positioning pilot signals across $L$ slots to a \ac{ue} equipped with $M_{\text{U}}$ antennas, who uses the received signals to positioning itself, referred to as \ac{bp}. Meanwhile, the \ac{bs} acts as a monostatic radar with $M_{\text{R}}$ colocated receive antennas, sensing the environments by receiving echoes from passive targets\footnote{To be noted, a passive target in \ac{ms} creates a multipath in \ac{bp}.} and the \ac{ue}, then estimating their positions, referred to as \ac{ms}. 

Let $N$ denote the number of \ac{ofdm} pilot symbols in each slot. The transmit signal associated with the $n$-th symbol in the $l$-th slot over the $m$-th subcarrier is given by
\begin{equation}\label{tran-sig}
\boldsymbol{x}_{l,n,m} = \boldsymbol{w}_{l}s_{n,m},
\end{equation}
 where $\boldsymbol{w}_{l}\in \mathbb{C}^{M_{\text{T}}  }$ is the beamformer for the $l$-th slot, and $s_{n,m}$ is the unit-modulus pilot symbol over the $m$-th subcarrier of the $n$-th symbol. 
 
 \subsubsection{Receive Signal at \ac{bp}}
 The signal received at the \ac{ue} is
\begin{equation}\label{re-sig-bi}
\overline{\boldsymbol{y}}_{l,n,m} = \overline{\boldsymbol{H}}_m\boldsymbol{x}_{l,n,m} + \overline{\boldsymbol{z}}_{l,n,m},
\end{equation}
where $\overline{\boldsymbol{H}}_m \in \mathbb{C}^{M_{\text{U}}\times M_{\text{T}}}$ is the channel between the \ac{bs} and the \ac{ue} over the $m$-th subcarrier, given by
\begin{equation}\label{bi-chan}
\overline{\boldsymbol{H}}_m = \sum_{k = 0}^{K} \alpha_{k}e^{-\jmath 2 \pi m \Delta f \tau_k} \boldsymbol{a}_{\text{U}}\left(\theta_{\text{U},k}\right)  \boldsymbol{a}_{\text{T}}^{\mathsf{H}}\left(\theta_{\text{B},k}\right),
\end{equation}
and $\overline{\boldsymbol{z}}_{l,n,m}\sim  \mathcal{CN}(\boldsymbol{0},\sigma^2 \boldsymbol{I}_{M_{\text{U}}})$ is the \ac{awgn} at the \ac{ue} receiver. Here, $\sigma^2 = F N_0 \Delta f$ is the noise power with $F$, $N_0$, and $\Delta f$ being the noise figure, single-side \ac{psd}, and subcarrier spacing, respectively, $K$ denotes the number of targets, and $\alpha_k$, $\tau_k$, $\theta_{\text{U},k}$, and $\theta_{\text{B},k}$ are the complex channel gain, delay, \ac{aoa}, and \ac{aod}, respectively, associated with the $k$-th path\footnote{For notational convenience, the \ac{los} path of the channel is indexed by $k=0$. Specifically, $\theta_{\text{U},0}$ and $\theta_{\text{B},0}$ denote the \ac{aoa} and \ac{aod} with respect to the \ac{bs} and the \ac{ue}, respectively.}. Finally, $\boldsymbol{a}_{\text{T}}\left(\theta\right) \in \mathbb{C}^{M_{\text{T}} }$ and $\boldsymbol{a}_{\text{U}}\left(\theta\right) \in \mathbb{C}^{M_{\text{U}} }$ are the steering vectors at the \ac{bs} (transmitter side) and the \ac{ue}, respectively.

\subsubsection{Receive Signal at \ac{ms}}
Similarly, the signal received at the \ac{bs} receiver is 
\begin{equation}\label{re-sig-mono}
\underline{\boldsymbol{y}}_{l,n,m} = \underline{\boldsymbol{H}}_m\boldsymbol{x}_{l,n,m} + \underline{\boldsymbol{z}}_{l,n,m},
\end{equation}
where $\underline{\boldsymbol{H}}_m \in \mathbb{C}^{M_{\text{R}}\times M_{\text{T}}}$ is the round-trip channel between the \ac{bs} and the passive targets (including the \ac{ue}) over the $m$-th subcarrier, given by
\begin{equation}\label{mono-chan}
\underline{\boldsymbol{H}}_m = \sum_{k = 0}^{K} \beta_{k}e^{-\jmath 2 \pi m \Delta f \epsilon_k} \boldsymbol{a}_{\text{R}}\left(\theta_{\text{B},k}\right)  \boldsymbol{a}_{\text{T}}^{\mathsf{H}}\left(\theta_{\text{B},k}\right),    
\end{equation}
and $\underline{\boldsymbol{z}}_{l,n,m}\sim  \mathcal{CN}(\boldsymbol{0},\sigma^2 \boldsymbol{I}_{M_{\text{R}}})$ is the \ac{awgn} at the \ac{bs} receiver. Here, $\beta_k$ and $\epsilon_k$ represent the complex channel gain and delay, respectively, associated with the $k$-th target\footnote{Note that the \ac{ue} is also an target (indexed by $k=0$) in the \ac{ms} scenario.}, while $\boldsymbol{a}_{\text{R}}\left(\theta\right) \in \mathbb{C}^{M_{\text{R}} }$ is the receiver-side steering vector at the \ac{bs}.

\subsection{CRB-Based Performance Metric}
For both \ac{bp} and \ac{ms}, we consider a two-stage positioning process, where the channel domain parameters are estimated in the first stage, and the position domain parameters are inferred from the channel domain parameters in the second stage.

\subsubsection{Performance Metric of \ac{bp}}
In \ac{bp}, the channel domain parameters are collected by $\overline{\boldsymbol{\xi}} = [\boldsymbol{\theta}_{\text{B}}^{\mathsf{T}},\boldsymbol{\theta}_{\text{U}}^{\mathsf{T}},\boldsymbol{\tau}^{\mathsf{T}},\boldsymbol{\alpha}_{\text{R}}^{\mathsf{T}},\boldsymbol{\alpha}_{\text{I}}^{\mathsf{T}}]^{\mathsf{T}} \in \mathbb{R}^{(5K+5) }$, where $\boldsymbol{\theta}_{\text{B}} = [{\theta}_{\text{B}, 0}, \ldots, {\theta}_{\text{B}, K}]^{\mathsf{T}} \in \mathbb{R}^{(K+1) }$ is the collection of \acp{aod}, $\boldsymbol{\theta}_{\text{U}} = [{\theta}_{\text{U}, 0}, \ldots, {\theta}_{\text{U}, K}]^{\mathsf{T}} \in \mathbb{R}^{(K+1) }$ is the collection of \acp{aoa}, $\boldsymbol{\tau} = [{\tau}_{0}, \ldots, {\tau}_{K}]^{\mathsf{T}} \in \mathbb{R}^{(K+1) }$ represents the delays, and $\boldsymbol{\alpha}_{\text{R}} = [\Re\{{\alpha}_{0}\}, \ldots, \Re\{{\alpha}_{K}\}]^{\mathsf{T}} \in \mathbb{R}^{(K+1) }$ and $\boldsymbol{\alpha}_{\text{I}} = [\Im\{{\alpha}_{0}\}, \ldots, \Im\{{\alpha}_{K}\}]^{\mathsf{T}} \in \mathbb{R}^{(K+1) }$ are the collections of the real and imaginary parts of the complex channel gains, respectively. Using the Slepian-Bangs formula\cite{furkan2022tvt}, the element at the $i$-th row and $j$-th column of the channel-domain \ac{fim} $\boldsymbol{I}_{\text{c}}(\overline{\boldsymbol{\xi}})$ is derived as
\begin{eqnarray}\label{bi-chan-fim}
\begin{aligned}
\left[\boldsymbol{I}_{\text{c}}\left(\overline{\boldsymbol{\xi}}\right)\right]_{i,j} &= \frac{2}{\sigma^2}\sum_{l=1}^{L}\sum_{n=1}^{N}
\sum_{m=1}^{M}\Re \left\{\frac{\partial \overline{\boldsymbol{\mu}}_{l,n,m}^{\mathsf{H}}}{\partial \left[\overline{\boldsymbol{\xi}}\right]_{i}}\frac{\partial \overline{\boldsymbol{\mu}}_{l,n,m}}{\partial \left[\overline{\boldsymbol{\xi}}\right]_{j}}\right\}\\
&= \frac{2N}{\sigma^2}\sum_{m=1}^{M}\Re \left\{\text{tr}\left(\frac{\partial \overline{\boldsymbol{H}}_{m}}{\partial \left[\overline{\boldsymbol{\xi}}\right]_{j}} \boldsymbol{W}\boldsymbol{W}^{\mathsf{H}}\frac{\partial \overline{\boldsymbol{H}}_{m}^{\mathsf{H}}}{\partial \left[\overline{\boldsymbol{\xi}}\right]_{i}}\right)\right\},
\end{aligned}
\end{eqnarray}
where $\overline{\boldsymbol{\mu}}_{l,n,m} = \overline{\boldsymbol{H}}_m\boldsymbol{x}_{l,n,m}$ denotes the noise-free observation from \eqref{re-sig-bi} and $\boldsymbol{W} = [\boldsymbol{w}_1,\ldots,\boldsymbol{w}_L] \in \mathbb{C}^{M_{\text{B}} \times L}$ collects $L$ beamformers.

The position-domain parameters are collected in $\overline{\boldsymbol{\eta}} = [\boldsymbol{p}_{\text{U}}^{\mathsf{T}},\phi,\boldsymbol{p}_{1}^{\mathsf{T}},\ldots,\boldsymbol{p}_{K}^{\mathsf{T}},\Delta t, \boldsymbol{\alpha}_{\text{R}}^{\mathsf{T}},\boldsymbol{\alpha}_{\text{I}}^{\mathsf{T}}]^{\mathsf{T}} \in \mathbb{R}^{(4K+6) }$, where $\boldsymbol{p}_{\text{U}} \in \mathbb{R}^{2  }$ represents the position of the \ac{ue}, and $\boldsymbol{p}_{k} \in \mathbb{R}^{2  }$ represents the position of the $k$-th target. The variable $\phi$ denotes the relative orientation of the \ac{bs} (in the \ac{ue}'s local coordinate system), while $\Delta t$ characterizes the clock bias that reflects the asynchronism between the \ac{bs} and \ac{ue} in the bistatic setting. Note that the nuisance parameters $\boldsymbol{\alpha}_{\text{R}}$ and $\boldsymbol{\alpha}_{\text{I}}$ from the channel-domain parameter $\overline{\boldsymbol{\xi}}$ remain part of the position-domain parameter $\overline{\boldsymbol{\eta}}$, as they do not contribute useful information for position estimation. Using the channel-domain \ac{fim}, the position-domain \ac{fim} $\boldsymbol{I}_{\text{p}}(\overline{\boldsymbol{\eta}})$ is computed as follows
\begin{equation}\label{bi-pos-fim}
\boldsymbol{I}_{\text{p}}\left(\overline{\boldsymbol{\eta}}\right) = \overline{\boldsymbol{J}}^{\mathsf{T}} \boldsymbol{I}_{\text{c}}\left(\overline{\boldsymbol{\xi}}\right)   \overline{\boldsymbol{J}},
\end{equation}
where $\overline{\boldsymbol{J}} \in  \mathbb{R}^{(5K+5)\times (4K+6)}$ is the Jacobian matrix, with the element in the $i$-th row and $j$-th column given by $[\overline{\boldsymbol{J}}]_{i,j} = \partial [\overline{\boldsymbol{\xi}}]_{i} / \partial [\overline{\boldsymbol{\eta}}]_{j}$. The \ac{crb} is used to quantify the \ac{bp} accuracy concerning $\boldsymbol{p}_{\text{U}}$, providing a lower bound on the sum of the variances for estimating $\boldsymbol{p}_{\text{U}}$, and is expressed as
\begin{equation}\label{crb-pos-bi}
\overline{\text{CRB}}\left(\boldsymbol{p}_{\text{U}}\right) = \text{tr}\left(\left[\boldsymbol{I}_{\text{p}}\left(\overline{\boldsymbol{\eta}}\right)^{-1}\right]_{1 : 2, 1 : 2}\right).    
\end{equation}

\subsubsection{Performance Metric of \ac{ms}}
Following similar steps, the position-domain \ac{fim} for \ac{ms} is given by
\begin{equation}\label{mono-pos-fim}
\boldsymbol{I}_{\text{p}}\left(\underline{\boldsymbol{\eta}}\right) = \underline{\boldsymbol{J}}^{\mathsf{T}} \boldsymbol{I}_{\text{c}}\left(\underline{\boldsymbol{\xi}}\right)   \underline{\boldsymbol{J}},
\end{equation}
where $\underline{\boldsymbol{\xi}} = [\boldsymbol{\theta}_{\text{B}}^{\mathsf{T}},\boldsymbol{\kappa}^{\mathsf{T}},\boldsymbol{\beta}_{\text{R}}^{\mathsf{T}},\boldsymbol{\beta}_{\text{I}}^{\mathsf{T}}]^{\mathsf{T}} \in \mathbb{R}^{(4K+4) }$ and $\underline{\boldsymbol{\eta}} = [\boldsymbol{p}_{\text{U}}^{\mathsf{T}},\boldsymbol{p}_{1}^{\mathsf{T}},\ldots,\boldsymbol{p}_{K}^{\mathsf{T}},\boldsymbol{\beta}_{\text{R}}^{\mathsf{T}},\boldsymbol{\beta}_{\text{I}}^{\mathsf{T}}]^{\mathsf{T}} \in \mathbb{R}^{(4K+4) }$ are the channel-domain and position-domain parameters, respectively. Here, $\boldsymbol{\kappa} = [{\kappa}_{0}, \ldots, {\kappa}_{K}]^{\mathsf{T}} \in \mathbb{R}^{(K+1) }$ represents the delay measurements, while $\boldsymbol{\beta}_{\text{R}} = [\Re\{{\beta}_{0}\}, \ldots, \Re\{{\beta}_{K}\}]^{\mathsf{T}} \in \mathbb{R}^{(K+1) }$ and $\boldsymbol{\beta}_{\text{I}} = [\Im\{{\beta}_{0}\}, \ldots, \Im\{{\beta}_{K}\}]^{\mathsf{T}} \in \mathbb{R}^{(K+1) }$ represent the real and imaginary parts of the complex channel gains, respectively. 
The \ac{crb} for \ac{ms}, concerning the passive targets (as well as the \ac{ue}), provides a lower bound on the sum variance for estimating $\boldsymbol{p} = [\boldsymbol{p}_{\text{U}}^{\mathsf{T}},\boldsymbol{p}_{1}^{\mathsf{T}},\ldots,\boldsymbol{p}_{K}^{\mathsf{T}}]^{\mathsf{T}} \in \mathbb{R}^{(2K+2) }$ at the \ac{bs}, and is given by
\begin{equation}\label{crb-pos-mono}
\underline{\text{CRB}}\left(\boldsymbol{p}\right) = \text{tr}\left(\left[\boldsymbol{I}_{\text{p}}\left(\underline{\boldsymbol{\eta}}\right)^{-1}\right]_{1 : 2K + 2, 1 : 2K + 2}\right).    
\end{equation}

\subsection{Problem Formulation}
We observe that both $\text{CRB}(\boldsymbol{p}_{\text{U}})$ and $\text{CRB}(\boldsymbol{p})$ are functions of $\boldsymbol{W}$, which can be optimized by designing the beamformers $\boldsymbol{W}$\cite{furkan2022tvt,henk2022jstsp}. However, due to the different objectives, a performance tradeoff between \ac{bp} and \ac{ms} emerges. Specifically, this bistatic-monostatic performance tradeoff is characterized by a \ac{moo} problem \cite{ehrgott2005multicriteria}, expressed as
\begin{subequations}\label{moo-prob}
\begin{align}
\mathop {\min }\limits_{\boldsymbol{W}} \;\;\; &\left[\overline{\text{CRB}}\left(\boldsymbol{p}_{\text{U}}\right), \underline{\text{CRB}}\left(\boldsymbol{p}\right)\right]\label{moo-prob-obj}\\
{\rm{s.t.}}\;\;\;
&  \text{tr}\left(\boldsymbol{W}\boldsymbol{W}^{\mathsf{H}}\right) \le P/M, \label{moo-prob-power}
\end{align}
\end{subequations}
where $P$ is the power budget\footnote{Without loss of generality, the right-hand side of \eqref{moo-prob-power} is set as $P/M$ such that the total transmit power over $M$ subcarriers is $P$.}. Note that the optimal solution to \eqref{moo-prob} represents the Pareto boundary of $[\overline{\text{CRB}}(\boldsymbol{p}_{\text{U}}), \underline{\text{CRB}}(\boldsymbol{p})]$, which is challenging to find due to the \ac{moo} nature. Additionally, neither $\overline{\text{CRB}}(\boldsymbol{p}_{\text{U}})$ nor $\underline{\text{CRB}}(\boldsymbol{p})$ is convex with respect to $\boldsymbol{W}$, further complicating the problem. 

\section{Tradeoff Between \ac{bp} and \ac{ms}}
\subsection{Weighted-Sum CRB Optimization}
To solve \eqref{moo-prob} and characterize the performance tradeoff, we first employ the weighted-sum approach, a classical method capable of obtaining the weak Pareto boundary of \ac{moo} problems\cite{ehrgott2005multicriteria}. Specifically, \eqref{moo-prob} is reformulated as 
\begin{subequations}\label{moo-prob-weight}
\begin{align}
\mathop {\min }\limits_{\boldsymbol{W}} \;\;\; &\rho \overline{\text{CRB}}\left(\boldsymbol{p}_{\text{U}}\right) + \left(1 - \rho\right)\underline{\text{CRB}}\left(\boldsymbol{p}\right)
\label{moo-prob-weight-obj}\\
{\rm{s.t.}}\;\;\;
&  \text{tr}\left(\boldsymbol{W}\boldsymbol{W}^{\mathsf{H}}\right) \le P/M, \label{moo-prob-weight-power}
\end{align}
\end{subequations}
where $\rho \in [0, 1]$ is a constant that adjusts the priority between \ac{bp} and \ac{ms}, determined by the specific application scenario and \ac{qos} requirements.

Problem \eqref{moo-prob-weight} remains challenging to solve due to its non-convexity. By defining $\boldsymbol{V} = \boldsymbol{W}\boldsymbol{W}^{\mathsf{H}}$, we lift \eqref{moo-prob-weight} into a relaxed form (by omitting the constraint $\text{rank}(\boldsymbol{V}) = L$) as
\begin{subequations}\label{moo-prob-weight-sdr}
\begin{align}
\mathop {\min }\limits_{\boldsymbol{V}} \;\;\; &\rho \overline{\text{CRB}}\left(\boldsymbol{p}_{\text{U}}\right) + \left(1 - \rho\right)\underline{\text{CRB}}\left(\boldsymbol{p}\right)
\label{moo-prob-weight-sdr-obj}\\
{\rm{s.t.}}\;\;\;
&  \text{tr}\left(\boldsymbol{V}\right) \le P/M, \quad \boldsymbol{V} \succeq \mathbf{0}. \label{moo-prob-weight-sdr-power}
\end{align}
\end{subequations}
Next, note that the matrices on the right-hand sides of \eqref{crb-pos-bi} and \eqref{crb-pos-mono} can be reformulated as\cite{henk2019twc}
\begin{subequations}\label{efim}
\begin{align}
&\left[\boldsymbol{I}_{\text{p}}\left(\overline{\boldsymbol{\eta}}\right)^{-1}\right]_{1 : 2, 1 : 2} = \left[\overline{\boldsymbol{F}} - \overline{\boldsymbol{G}}\overline{\boldsymbol{Z}}^{-1}\overline{\boldsymbol{G}}^{\mathsf{T}}\right]^{-1}, \\
&\left[\boldsymbol{I}_{\text{p}}\left(\underline{\boldsymbol{\eta}}\right)^{-1}\right]_{1 : 2K + 2, 1 : 2K + 2} = \left[\underline{\boldsymbol{F}} - \underline{\boldsymbol{G}}\underline{\boldsymbol{Z}}^{-1}\underline{\boldsymbol{G}}^{\mathsf{T}}\right]^{-1},
\end{align}
\end{subequations}
where $\overline{\boldsymbol{F}} = [\boldsymbol{I}_{\text{p}}(\overline{\boldsymbol{\eta}})]_{1 : 2, 1 : 2}$, $\overline{\boldsymbol{G}} = [\boldsymbol{I}_{\text{p}}(\overline{\boldsymbol{\eta}})]_{1 : 2, 3 : 4K + 6}$, $\overline{\boldsymbol{Z}} = [\boldsymbol{I}_{\text{p}}(\overline{\boldsymbol{\eta}})]_{3 : 4K + 6, 3 : 4K + 6}$, $\underline{\boldsymbol{F}} = [\boldsymbol{I}_{\text{p}}(\underline{\boldsymbol{\eta}})]_{1 : 2K + 2, 1 : 2K + 2}$, $\underline{\boldsymbol{G}} = [\boldsymbol{I}_{\text{p}}(\underline{\boldsymbol{\eta}})]_{1 : 2K + 2, 2K + 3 : 4K + 4}$, and $\underline{\boldsymbol{Z}} = [\boldsymbol{I}_{\text{p}}(\underline{\boldsymbol{\eta}})]_{2K + 3 : 4K + 4, 2K + 3 : 4K + 4}$. 
By introducing auxiliary variables $\overline{\boldsymbol{U}}\in \mathbb{R}^{2 \times 2}$ and $\underline{\boldsymbol{U}}\in \mathbb{R}^{(2K+2)\times (2K+2)}$, \eqref{moo-prob-weight-sdr} can be reformulated into an equivalent form as
\begin{subequations}\label{moo-prob-weight-relax}
\begin{align}
\mathop {\min }\limits_{\boldsymbol{V},\overline{\boldsymbol{U}},\underline{\boldsymbol{U}}} \;\;\; &\rho \text{tr}\left(\overline{\boldsymbol{U}}^{-1}\right) + \left(1 - \rho\right)\text{tr}\left(\underline{\boldsymbol{U}}^{-1}\right)
\label{moo-prob-weight-relax-obj}\\
{\rm{s.t.}}\;\;\;
& 
\begin{bmatrix}
\overline{\boldsymbol{F}} - \overline{\boldsymbol{U}} & \overline{\boldsymbol{G}}\\
\overline{\boldsymbol{G}}^{\mathsf{T}} & \overline{\boldsymbol{Z}}
\end{bmatrix} \succeq \mathbf{0}, \quad
\begin{bmatrix}
\underline{\boldsymbol{F}} - \underline{\boldsymbol{U}} & \underline{\boldsymbol{G}}\\
\underline{\boldsymbol{G}}^{\mathsf{T}} & \underline{\boldsymbol{Z}}
\end{bmatrix} \succeq \mathbf{0},\label{moo-prob-weight-relax-lmi}\\
& \overline{\boldsymbol{U}} \succeq \mathbf{0}, \quad \underline{\boldsymbol{U}} \succeq \mathbf{0}, \label{moo-prob-weight-relax-psd}\\
&  \text{tr}\left(\boldsymbol{V}\right) \le P/M,  \quad \boldsymbol{V} \succeq \mathbf{0}. 
\end{align}
\end{subequations}

The above problem is a convex \ac{sdp} problem that can be efficiently solved using off-the-shelf optimization tools such as CVX. Once solved, the beamformers $\boldsymbol{W}$ can be recovered from $\boldsymbol{V}$ via matrix decomposition or a randomization procedure\cite{tom2010spm}.

\subsection{Weighted-Sum Mismatch Approaches} 
Inspired by the weighted waveform mismatch minimization approach commonly used in the \ac{isac} literature to balance sensing and communication performance \cite{fan2018tsp,youli2024twc}, we propose two alternative methods to effectively balance \ac{bp} and \ac{ms} by minimizing the weighted-sum mismatch of various metrics. Specifically, the optimal beamformers $\overline{\boldsymbol{W}}$ for \ac{bp} and $\underline{\boldsymbol{W}}$ for \ac{ms} are obtained by solving \eqref{moo-prob-weight-relax} for $\rho = 1$ and $\rho = 0$, respectively. The \emph{balanced} beamformers are then derived from these extremes using different strategies to weigh the mismatch: one approach minimizes the weighted-sum mismatch of the beamformers, while the other minimizes the weighted-sum mismatch of the variance matrices.

\subsubsection{Weighted-Sum Mismatch of Beamformers}
Upon obtaining $\overline{\boldsymbol{W}}$ and $\underline{\boldsymbol{W}}$, we formulate the following optimization problem to minimize the weighted-sum mismatch of beamformers
\vspace{-5mm}
\begin{subequations}\label{moo-bf-approx}
\begin{align}
\mathop {\min }\limits_{\boldsymbol{W}} \;\;\; 
&\rho \left\|\boldsymbol{W} - \overline{\boldsymbol{W}} \right\|_{\text{F}}^{2} + \left(1 - \rho\right)\left\|\boldsymbol{W} - \underline{\boldsymbol{W}} \right\|_{\text{F}}^{2} \label{moo-bf-approx-obj}\\
{\rm{s.t.}}\;\;\; 
&\text{tr}\left(\boldsymbol{W}\boldsymbol{W}^{\mathsf{H}}\right) = P/M, \label{moo-bf-approx-power}
\end{align}
\end{subequations}
where \eqref{moo-bf-approx-power} represents the full power transmission constraint. Problem \eqref{moo-bf-approx} can be equivalently reformulated as
\begin{subequations}\label{moo-bf-approx-ref}
\begin{align}
\mathop {\min }\limits_{\boldsymbol{W}} \;\;\; 
&\left\|\boldsymbol{A}\boldsymbol{W} - \boldsymbol{B} \right\|_{\text{F}}^{2}  \label{moo-bf-approx-ref-obj}\\
{\rm{s.t.}}\;\;\; 
&\text{tr}\left(\boldsymbol{W}\boldsymbol{W}^{\mathsf{H}}\right) = P/M, \label{moo-bf-approx-ref-power}
\end{align}
\end{subequations}
where $\boldsymbol{A} = [\sqrt{\rho}\boldsymbol{E}_{M_{\text{T}}};\sqrt{1-\rho}\boldsymbol{E}_{M_{\text{T}}}]$ and $\boldsymbol{B} = [\sqrt{\rho}\overline{\boldsymbol{W}};\sqrt{1-\rho}\underline{\boldsymbol{W}}]$. 

This problem is non-convex due to the equality constraint in \eqref{moo-bf-approx-ref-power}. Notably, $\boldsymbol{A}^{\mathsf{H}}\boldsymbol{A} = \boldsymbol{E}_{M_{\text{T}}}$. By defining $\boldsymbol{\Xi} = \boldsymbol{A}^{\mathsf{H}}\boldsymbol{B}$ and $\tilde{\boldsymbol{W}}_l = \tilde{\boldsymbol{w}}_l\tilde{\boldsymbol{w}}_l^{\mathsf{H}}$, where $\tilde{\boldsymbol{w}}_l = [1,\boldsymbol{w}^{\mathsf{T}}]^{\mathsf{T}}$, we can relax the constraint $\text{rank}(\tilde{\boldsymbol{W}}_l) = 1$. This enables us to lift and reformulate \eqref{moo-bf-approx-ref} into
\begin{subequations}\label{moo-bf-approx-ref-sdp}
\begin{align}
\mathop {\min }\limits_{\tilde{\boldsymbol{W}_l}} \;\;\; 
&\sum_{l=1}^{L} \text{tr}\left( 
\begin{bmatrix}
0 & -\boldsymbol{\Xi}\left(:,l\right)^{\mathsf{H}}\\
-\boldsymbol{\Xi}\left(:,l\right) & \boldsymbol{E}_{M_{\text{T}}}
\end{bmatrix}
\tilde{\boldsymbol{W}_l}\right)  \label{moo-bf-approx-ref-sdp-obj}\\
{\rm{s.t.}}\;\;\; 
& \sum_{l=1}^{L} \text{tr}\left(\tilde{\boldsymbol{W}_l}\right) = P/M + L, \label{moo-bf-approx-ref-sdp-power} \\
&\left[\tilde{\boldsymbol{W}_l}\right]_{1,1} = 1, \quad \tilde{\boldsymbol{W}_l} \succeq \mathbf{0}, \quad l=1,\ldots,L.
\end{align}
\end{subequations}

Problem \eqref{moo-bf-approx-ref-sdp} is a convex \ac{sdp} problem. Although relaxed, it falls under the category of trust-region subproblems, which exhibit strong duality and are guaranteed to yield rank-one solutions \cite{fortin2004trust}. This implies that the optimal $\tilde{\boldsymbol{w}}_l$, and thus the optimal $\boldsymbol{w}_l$ for \eqref{moo-bf-approx}, can be recovered from the obtained $\tilde{\boldsymbol{W}_l}$.

\subsubsection{Weighted-Sum Mismatch of Variance Matrices}
We observe that the position-domain \ac{fim}, and consequently the \acp{crb} in \eqref{crb-pos-bi} and \eqref{crb-pos-mono}, are directly determined by $\boldsymbol{W}\boldsymbol{W}^{\mathsf{H}}$. This can be viewed as the \emph{variance matrices} of the transmit signal (ignoring scaling) and reflects the accumulated effect of different beamformers. Inspired by this observation, we propose minimizing the weighted sum mismatch of the variance matrices. The beamformers can then be recovered from the obtained variance matrix. Specifically, the optimization problem can be formulated as 
\begin{subequations}\label{moo-bf-approx-var}
\begin{align}
\mathop {\min }\limits_{\boldsymbol{V}} \;\;\; 
&\rho \left\|\boldsymbol{V} - \overline{\boldsymbol{V}} \right\|_{\text{F}}^{2} + \left(1 - \rho\right)\left\|\boldsymbol{V} - \underline{\boldsymbol{V}} \right\|_{\text{F}}^{2} \label{moo-bf-approx-var-obj}\\
{\rm{s.t.}}\;\;\; 
&\text{tr}\left(\boldsymbol{V}\right) = P/M, \label{moo-bf-approx-var-power}\\
&\boldsymbol{V} \succeq \boldsymbol{0},\quad \text{rank}\left(\boldsymbol{V}\right) = L, \label{moo-bf-approx-var-rank}
\end{align}
\end{subequations}
where $\overline{\boldsymbol{V}} = \overline{\boldsymbol{W}}\overline{\boldsymbol{W}}^{\mathsf{H}}$ and $\underline{\boldsymbol{V}} = \underline{\boldsymbol{W}}\underline{\boldsymbol{W}}^{\mathsf{H}}$.

Problem \eqref{moo-bf-approx-var} can be efficiently solved using the \ac{sdr} technique. Specifically, by temporarily dropping the rank constraint, \eqref{moo-bf-approx-var} becomes an \ac{sdp} problem, from which the optimal variance matrix $\boldsymbol{V}$ can be obtained. Subsequently, the beamformers $\boldsymbol{W}$ are retrieved using a randomization procedure\cite{tom2010spm}.

\subsection{Complexity Analysis}
According to\cite{furkan2022tvt}, the computational complexity of an \ac{sdp} problem is given by $\mathcal{O}(I^2 \sum_{j=1}^{J}d_j^2 + I \sum_{j=1}^{J}d_j^3)$, where $I$ and $J$ represent the numbers of optimization variables and \ac{lmi} constraints, respectively, and $d_j$ denotes the row/column size of the matrix associated with the $j$-th \ac{lmi}.
\begin{itemize}
    \item For \eqref{moo-prob-weight-relax}: $I = M_{\text{T}}^2 + (2K+2)^2 + 4$, $J = 5$, $d_1 = 4K+6$, $d_2 = 4K+4$, $d_3 = 2$, $d_4 = 2K+2$, and $d_5 = M_{\text{T}}$.
    \item For \eqref{moo-bf-approx-ref-sdp}: $I = M_{\text{T}}^2 L$, $J = L$, and $d_j = M_{\text{T}}$.
    \item For \eqref{moo-bf-approx-var}: $I = M_{\text{T}}^2$, $J = 1$, and $d_j = M_{\text{T}}$. 
\end{itemize}
When $K \ll M_{\text{T}}$ (e.g., in scenarios with channel sparsity), the computational complexity of solving each problem can be approximated as $\mathcal{O}(M_{\text{T}}^6)$. Note that, although \eqref{moo-bf-approx-ref-sdp} and \eqref{moo-bf-approx-var} are subject to roughly the same order of complexity, their actual running times in simulations are usually significantly lower than that of \eqref{moo-prob-weight-relax}.



\begin{figure*}[t]
\begin{minipage}[b]{0.99\linewidth}
\begin{center}
\end{center}
\centering
\centerline{\includegraphics[width=1\linewidth]{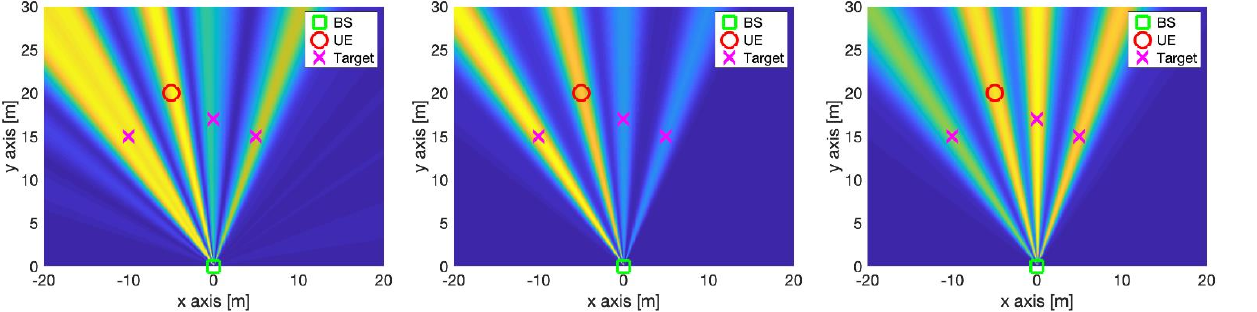} }
\small
\centerline{
(a) \ac{bp}-FDB 
\hspace{4.0cm} 
(b) \ac{bp}-CPA 
\hspace{4.0cm} 
(c) \ac{bp}-APA
}
\normalsize
\end{minipage}

\begin{minipage}[b]{0.99\linewidth}
\centering
\centerline{\includegraphics[width=1\linewidth]{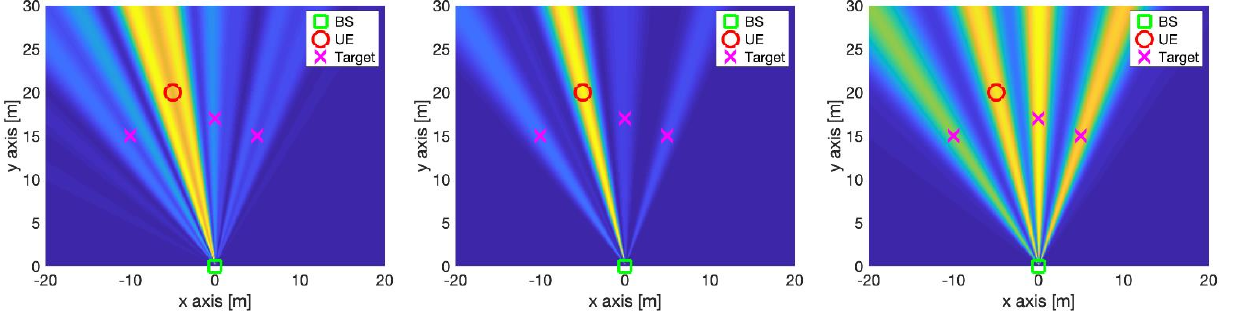}}
\small
\centerline{
(d) \ac{ms}-FDB
\hspace{4.0cm} 
(e) \ac{ms}-CPA
\hspace{4.0cm} 
(f) \ac{ms}-APA
}
\normalsize
\end{minipage}

\caption{Beampatterns: (a) \ac{bp}-FDB; (b) \ac{bp}-CPA; (c) \ac{bp}-APA; (d) \ac{ms}-FDB; (e) \ac{ms}-CPA; (f) \ac{ms}-APA.}
\vspace{-3mm}
\label{bp}
\end{figure*}
\section{Numerical Results}
\subsection{Scenarios}
Unless stated otherwise, the simulation parameters are as follows: The \ac{bs} is equipped with $M_{\text{T}} = 16$ transmit antennas and $M_{\text{R}} = 16$ colocated receive antennas, positioned at $\boldsymbol{p}_{\text{B}} = [0\text{ m}, 0\text{ m}]^{\mathsf{T}}$. The \ac{ue}, equipped with $M_\text{U} = 16$ antennas, is located at $\boldsymbol{p}_{\text{U}} = [-5\text{ m}, 20\text{ m}]^{\mathsf{T}}$. Additionally, there are $K=3$ targets, positioned at $\boldsymbol{p}_1 = [-10\text{ m}, 15\text{ m}]^{\mathsf{T}}$, $\boldsymbol{p}_2 = [5\text{ m}, 15\text{ m}]^{\mathsf{T}}$, and $\boldsymbol{p}_3 = [0\text{ m}, 17\text{ m}]^{\mathsf{T}}$, respectively.
The transmit power is set to $P = -20\text{ dBm}$, with a carrier frequency of $f_c = 28 \text{ GHz}$ and a bandwidth of $W = 120 \text{ MHz}$. The number of subcarriers is $M = 1024$, the noise figure is $F = 10 \text{ dB}$, and the noise \ac{psd} is $N_0 = -173.855 \text{ dBm/Hz}$. The system simulates $L = 16$ slots, each with $N = 100$ pilots, and the clock bias is $\Delta t = 1\;\mu\text{s}$. The \ac{ue}'s relative orientation is $\phi = (110/180)\pi$.
The channel gains are generated using a standard free-space path loss model. For the $k$-th path, the phases $\overline{\zeta}_k$ (for \ac{bp}) and $\underline{\zeta}_k$ (for \ac{ms}) are uniformly distributed over $[-\pi, \pi]$. In \ac{bp}, the \ac{los} channel gain is $\alpha_0 = e^{\jmath \overline{\zeta}_0} \lambda/(4 \pi \left\|\boldsymbol{p}_{\text{B}} - \boldsymbol{p}_{\text{U}}\right\|)$, while the \ac{nlos} channel gain is $\alpha_k = \overline{\sigma}_{\text{RCS}}e^{\jmath \overline{\zeta}_k} \lambda/((4 \pi)^{3/2} \left\|\boldsymbol{p}_{\text{U}} - \boldsymbol{p}_k\right\| \left\|\boldsymbol{p}_k - \boldsymbol{p}_{\text{B}}\right\|)$.
Here, $\overline{\sigma}_{\text{RCS},k}$ (for \ac{bp}) and $\underline{\sigma}_{\text{RCS},k}$ (for \ac{ms}) represent the \ac{rcs} of the $k$-th target while $\lambda$ is the wavelength. Specifically, $\underline{\sigma}_{\text{RCS},0} = 10 \text{ m}^2$, while $\overline{\sigma}_{\text{RCS},k} = \underline{\sigma}_{\text{RCS},k} = 100 \text{ m}^2$ ($k=1,\ldots,K$).
\subsection{Compared Schemes}
We characterize the performance tradeoff boundary between \ac{bp} and \ac{ms} through three categories of approaches, which are detailed in the following. 
\subsubsection{FDB} The first category focuses on solving \ac{fdb} optimization problems, represented by \eqref{moo-prob-weight}, \eqref{moo-bf-approx}, and \eqref{moo-bf-approx-var}. These are further categorized as \ac{fdb}-weighted \ac{crb} (\ac{fdb}-WCRB), \ac{fdb}-weighted beamformer (\ac{fdb}-WBF), and \ac{fdb}-weighted variance matrix (\ac{fdb}-WVM), respectively.

\subsubsection{CPA} As a low-complexity alternative, we introduce a \ac{cpa} approach. The core idea is that the optimal variance matrix minimizing the \ac{crb} can be expressed as $\boldsymbol{V} = \boldsymbol{U}\boldsymbol{\Lambda}\boldsymbol{U}^{\mathsf{H}}$, where $\boldsymbol{\Lambda} \in \mathbb{C}^{\left(2K + 2\right) \times \left(2K + 2\right)}$ and $\boldsymbol{U} = [\boldsymbol{a}_{\text{T}}\left(\theta_{\text{B},0}\right), \ldots, \boldsymbol{a}_{\text{T}}\left(\theta_{\text{B},K}\right), \dot{\boldsymbol{a}}_{\text{T}}\left(\theta_{\text{B},0}\right), \ldots, \dot{\boldsymbol{a}}_{\text{T}}\left(\theta_{\text{B},K}\right)] \in \mathbb{C}^{M_{\text{T}} \times \left(2K + 2\right)}$, with $\partial \boldsymbol{a}_{\text{T}}\left(\theta_{\text{B},k}\right)/\partial \theta$\cite{furkan2022tvt,henk2022jstsp}. 
Moreover, by restricting $\boldsymbol{\Lambda}$ to be diagonal, this approach simplifies to a lower-dimensional, low-complexity power allocation problem over the predetermined codebook $\boldsymbol{U}$. In other words, the optimization problem in \eqref{moo-prob-weight} can be reformulated as a low-complexity power allocation task, and the solutions for \eqref{moo-bf-approx} and \eqref{moo-bf-approx-var} can be derived using the variance matrix obtained from solving \eqref{moo-prob-weight}. These methods are further categorized as \ac{cpa} weighted \ac{crb} (\ac{cpa}-WCRB), \ac{cpa} weighted beamformer (\ac{cpa}-WBF), and \ac{cpa} weighted variance matrix (\ac{cpa}-WVM), respectively.

\subsubsection{APA} Lastly, a simple baseline is the \ac{apa} across the given codebook $\boldsymbol{U}$, where no distinction or tradeoff is made between \ac{bp} and \ac{ms}.


\subsection{Results and Discussion}
\subsubsection{Beampatterns}
In Fig. \ref{bp}, we show the optimized beampatterns. The top row illustrates the case of \ac{bp} ($\rho = 1$), while the bottom row corresponds to \ac{ms} ($\rho = 0$). For each row, the results for the \ac{fdb}, \ac{cpa}, and \ac{apa} schemes are displayed from left to right, respectively. We observe that under \ac{fdb} and \ac{cpa}, the beampatterns exhibit distinct characteristics between \ac{bp} and \ac{ms}, whereas under \ac{apa}, there is no difference. This is because adaptive design (tailored for different $\rho$) is applied in the \ac{fdb} and \ac{cpa} schemes but not the \ac{apa} case. 

Focusing on the cases under the \ac{fdb} and \ac{cpa} schemes, we observe that for \ac{bp}, strong beams are directed toward the \ac{ue} in both schemes, while the power allocated to beams directed at the targets varies. This is because the \ac{bs} acts as an essential anchor in \ac{bp}, whereas the targets serve as auxiliary anchors, with their positions being estimated simultaneously. Depending on the relative locations, different amounts of information are provided regarding the \ac{ue}'s position. Therefore, beams with varying power levels are allocated to illuminate the targets, maximizing the \ac{ue}'s positioning accuracy. In \ac{ms}, all targets and the \ac{ue} need to be accurately positioned for optimal performance. Since the \ac{ue} has the smallest \ac{rcs}, stronger beams are directed toward it under both the \ac{fdb} and \ac{cpa} schemes to ensure balanced positioning performance across all targets.

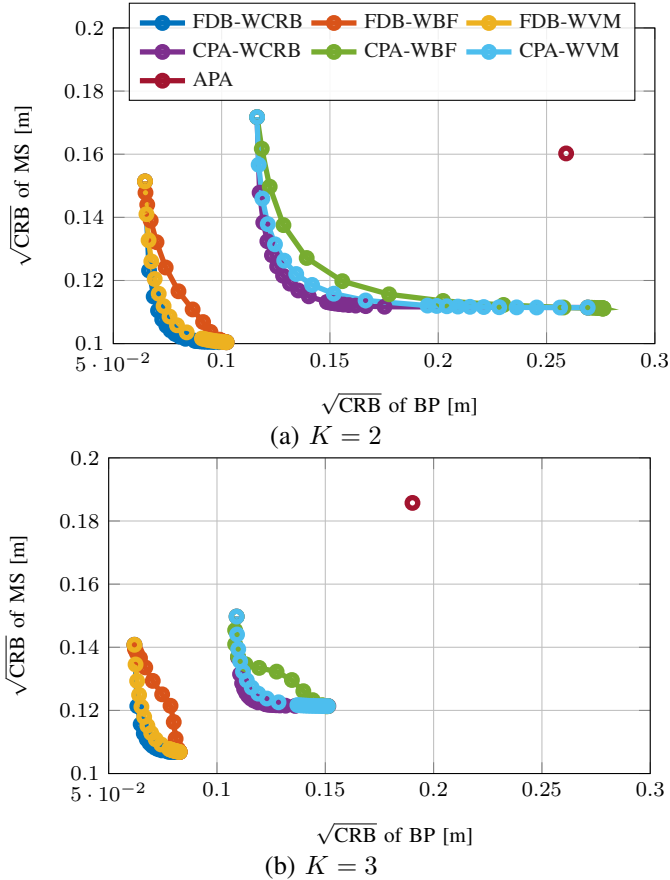
\begin{figure}[htb]
\centering
\begin{minipage}[b]{0.98\linewidth}
\vspace{-0.3cm}
  \centering
%
%
\definecolor{mycolor1}{rgb}{0.00000,0.44700,0.74100}%
\definecolor{mycolor2}{rgb}{0.85000,0.32500,0.09800}%
\definecolor{mycolor3}{rgb}{0.92900,0.69400,0.12500}%
\definecolor{mycolor4}{rgb}{0.49400,0.18400,0.55600}%
\definecolor{mycolor5}{rgb}{0.46600,0.67400,0.18800}%
\definecolor{mycolor6}{rgb}{0.30100,0.74500,0.93300}%
\definecolor{mycolor7}{rgb}{0.63500,0.07800,0.18400}%
\begin{tikzpicture}

\begin{axis}[%
width=72mm,
height=42mm,
at={(0in,0in)},
scale only axis,
xmin=0.05,
xmax=0.3,
xticklabel style = {font=\color{white!15!black},font=\footnotesize},
xlabel style={font=\footnotesize, xshift=2mm},
xlabel={$\sqrt{\text{CRB}}$ of \ac{bp} [m]},
ymin=0.1,
ymax=0.2,
yticklabel style = {font=\color{white!15!black},font=\footnotesize},
ylabel style={font=\footnotesize, yshift=0mm,font=\footnotesize},
ylabel={$\sqrt{\text{CRB}}$ of \ac{ms} [m]},
axis background/.style={fill=white},
xmajorgrids,
ymajorgrids,
legend style={font=\footnotesize, at={(0.964,1.08)}, anchor=north east, legend cell align=left, align=left, fill = white, fill opacity=0.9, legend columns = 3}
]
\addplot [color=mycolor1, line width=2.0pt, mark=o, mark options={solid, mycolor1}]
  table[row sep=crcr]{%
0.102312199479144	0.100444798205856\\
0.0987180867711027	0.10046154994497\\
0.0961637827339674	0.100498547221058\\
0.0942445558809087	0.100544722930025\\
0.0927351151381106	0.100595328424586\\
0.0915060023926092	0.100648158844642\\
0.0904805554642126	0.100701959527756\\
0.0895986394695089	0.100756629704012\\
0.0888325353539667	0.100811469784934\\
0.0881495278225621	0.100867030923972\\
0.0875345991561757	0.100923176229952\\
0.0833249368248501	0.101529503515772\\
0.0807355867528395	0.102199973670008\\
0.0785072410524487	0.103063038324872\\
0.076435403093412	0.104198542410541\\
0.0745082229676217	0.105654680762206\\
0.0725838406273655	0.107655576093996\\
0.0706522870750803	0.110489002401215\\
0.0686437809428402	0.114912078673458\\
0.0665060477740433	0.123181108218556\\
0.0646330552067132	0.151394172788542\\
};
\addlegendentry{FDB-WCRB}

\addplot [color=mycolor2, line width=2.0pt, mark=o, mark options={solid, mycolor2}]
  table[row sep=crcr]{%
0.102305801705171	0.100444739536694\\
0.10203590647049	0.100452430927664\\
0.101749567175178	0.100475758515583\\
0.101446954492221	0.10051507902302\\
0.101128430881588	0.100570721406709\\
0.100794558757621	0.10064298019685\\
0.100446105728243	0.100732107994685\\
0.100084045922666	0.100838306848118\\
0.0997095569064363	0.10096171873147\\
0.0993240110916147	0.101102415282766\\
0.0989289609660689	0.101260387076597\\
0.0948545381669485	0.103713982419628\\
0.0914486598933561	0.10683295774237\\
0.0866195178658088	0.110828528171583\\
0.0800858740758152	0.116578285358323\\
0.0741777213743385	0.124115747247893\\
0.069953459887994	0.132137976412452\\
0.0672588038166198	0.138999107026716\\
0.0656717463094579	0.144022879485251\\
0.0648663065691367	0.147808978643124\\
0.0646329670178888	0.151388739016414\\
};
\addlegendentry{FDB-WBF}

\addplot [color=mycolor3, line width=2.0pt, mark=o, mark options={solid, mycolor3}]
  table[row sep=crcr]{%
0.102305757811735	0.100444739189689\\
0.100846001073435	0.100524707445205\\
0.0994761330659173	0.100614095336542\\
0.0981903710714336	0.100713396517498\\
0.0969748492999696	0.10082013599071\\
0.0958294180069686	0.100936426195718\\
0.0947463934455345	0.101061432609622\\
0.0937194376987154	0.101194460218383\\
0.0927423378047688	0.101334552478407\\
0.0918154178605265	0.101483620051845\\
0.0909258660130834	0.101637048681092\\
0.0838667170410823	0.103564378496682\\
0.0794050222367728	0.105816192698075\\
0.0759668401035578	0.108575147862121\\
0.0732184208283679	0.111880162692155\\
0.0709692816983822	0.115805147987938\\
0.0691063545433356	0.120467799444158\\
0.0675593409242852	0.126034984001694\\
0.0662919152015156	0.132753428762916\\
0.0653033161272268	0.141010105392052\\
0.0646329675226152	0.151388737486307\\
};
\addlegendentry{FDB-WVM}

\addplot [color=mycolor4, line width=2.0pt, mark=o, mark options={solid, mycolor4}]
  table[row sep=crcr]{%
0.269059225361796	0.111419006185022\\
0.175206394850328	0.111681712927429\\
0.16653654891952	0.111873886442596\\
0.161887383965269	0.112045965243269\\
0.158712486011329	0.112209345467663\\
0.156272510402938	0.112369862502097\\
0.154278429315402	0.112529645961496\\
0.152584386640899	0.112689773571486\\
0.151099600511505	0.112851630756292\\
0.149781651621213	0.113014550833157\\
0.148586397142107	0.113179901160491\\
0.140254058308557	0.114979865630016\\
0.135381183187959	0.116839850109992\\
0.131549490423588	0.119000893233286\\
0.128362295748703	0.12151486436967\\
0.125644058690561	0.124447571196784\\
0.123237418070599	0.128010390414438\\
0.121120169234606	0.13243208083311\\
0.119229729248835	0.138375597037395\\
0.117535145321842	0.147822189607496\\
0.116372293821203	0.171751112962658\\
};
\addlegendentry{CPA-WCRB}

\addplot [color=mycolor5, line width=2.0pt, mark=o, mark options={solid, mycolor5}]
  table[row sep=crcr]{%
0.26899791245842	0.111419014616497\\
0.270627424789467	0.111370384108337\\
0.272082655073581	0.111325748123324\\
0.273346280304607	0.111285322601776\\
0.274401858848283	0.111249329902628\\
0.27523421981837	0.111217997312657\\
0.275829853064997	0.111191555204668\\
0.276177278555039	0.111170234832754\\
0.27626738882538	0.111154265759444\\
0.276093732374646	0.111143872931791\\
0.275652737366203	0.111139273425341\\
0.257833043446918	0.111446679583053\\
0.230136818858992	0.112261913046133\\
0.202235171895661	0.113523174308313\\
0.177302867780383	0.115618736118047\\
0.155720889708412	0.119782318555275\\
0.139275715475301	0.127128329543763\\
0.128545521649382	0.13754155115051\\
0.122199031965172	0.14971909626955\\
0.118545443186799	0.161733932498305\\
0.116372170805486	0.171750501142281\\
};
\addlegendentry{CPA-WBF}

\addplot [color=mycolor6, line width=2.0pt, mark=o, mark options={solid, mycolor6}]
  table[row sep=crcr]{%
0.268996640187718	0.111419006669817\\
0.256251348985263	0.111427850105224\\
0.245428840755588	0.111449257525738\\
0.236189408731031	0.111484353228927\\
0.228129980355737	0.111531537296652\\
0.221006602570311	0.111589940095817\\
0.214652095857728	0.111659280304317\\
0.209060333760445	0.111742788838246\\
0.203977641979526	0.111833982186078\\
0.199314889767589	0.111936190040475\\
0.19503637379253	0.112046977244701\\
0.166457127266483	0.113675453966284\\
0.151784996307547	0.115820300021682\\
0.141764486781028	0.118598470627342\\
0.134439504784687	0.122051702187214\\
0.12889053361665	0.126280569903266\\
0.124636646771169	0.131461389923187\\
0.121302753540121	0.137862318790144\\
0.118812062995827	0.145971492749337\\
0.117103100349872	0.156657863797792\\
0.116372294758149	0.171750600674566\\
};
\addlegendentry{CPA-WVM}

\addplot [color=mycolor7, line width=2.0pt, mark=o, mark options={solid, mycolor7}]
  table[row sep=crcr]{%
0.259035535380864	0.160245901782767\\
};
\addlegendentry{APA}

\end{axis}
\end{tikzpicture}%
    \vspace{-1.cm}
  \centerline{(a) $K=2$} \medskip
\end{minipage}
\hfill
\begin{minipage}[b]{0.98\linewidth}
\vspace{-0.3cm}
  \centering
%
%
\definecolor{mycolor1}{rgb}{0.00000,0.44700,0.74100}%
\definecolor{mycolor2}{rgb}{0.85000,0.32500,0.09800}%
\definecolor{mycolor3}{rgb}{0.92900,0.69400,0.12500}%
\definecolor{mycolor4}{rgb}{0.49400,0.18400,0.55600}%
\definecolor{mycolor5}{rgb}{0.46600,0.67400,0.18800}%
\definecolor{mycolor6}{rgb}{0.30100,0.74500,0.93300}%
\definecolor{mycolor7}{rgb}{0.63500,0.07800,0.18400}%
\begin{tikzpicture}

\begin{axis}[%
width=72mm,
height=42mm,
at={(0in,0in)},
scale only axis,
xmin=0.05,
xmax=0.3,
xticklabel style = {font=\color{white!15!black},font=\footnotesize},
xlabel style={font=\footnotesize, xshift=2mm},
xlabel={$\sqrt{\text{CRB}}$ of \ac{bp} [m]},
ymin=0.1,
ymax=0.2,
yticklabel style = {font=\color{white!15!black},font=\footnotesize},
ylabel style={font=\footnotesize, yshift=0mm},
ylabel={$\sqrt{\text{CRB}}$ of \ac{ms} [m]},
axis background/.style={fill=white},
xmajorgrids,
ymajorgrids,
legend style={font=\scriptsize, at={(1,1)}, anchor=north east, legend cell align=left, align=left, fill = white, fill opacity=0.9, legend columns = 3}
]
\addplot [color=mycolor1, line width=2.0pt, mark=o, mark options={solid, mycolor1}]
  table[row sep=crcr]{%
0.0827873031720473	0.106839978252764\\
0.0818947385382188	0.106841871111424\\
0.0810946444238936	0.106851017212691\\
0.0804055044780092	0.106864335053041\\
0.0797949868044433	0.106880862777541\\
0.0792462621964328	0.106900008260263\\
0.0787528510518854	0.106921188392837\\
0.078301897979017	0.106944173536233\\
0.0778876317776313	0.10696866809789\\
0.0775047115779013	0.106994476057877\\
0.0771487021778817	0.107021456670442\\
0.0744989378276438	0.107343488471397\\
0.0728382776934224	0.107709549410139\\
0.0714324009811981	0.108179279990219\\
0.0701310915395729	0.108801368056383\\
0.068857596743428	0.109652378417299\\
0.0675650693156507	0.110860217158332\\
0.0662144617155322	0.112669696506286\\
0.0647727293892009	0.115622544636856\\
0.0632136137929659	0.121363694629064\\
0.0618872332542553	0.140793205288059\\
};

\addplot [color=mycolor2, line width=2.0pt, mark=o, mark options={solid, mycolor2}]
  table[row sep=crcr]{%
0.0828147349533695	0.106838396480586\\
0.0827351984458316	0.106846508838233\\
0.0826541461967008	0.106871272447084\\
0.0825714410092159	0.106913351347382\\
0.0824869648558893	0.106973434760038\\
0.0824006204409915	0.107052238994223\\
0.0823123325421646	0.107150509234761\\
0.0822220497324462	0.107269021253377\\
0.0821297460867577	0.107408582614862\\
0.0820354232496091	0.107570033326742\\
0.0819391127565331	0.107754245424218\\
0.0809010553302306	0.111048885657095\\
0.0800407664599757	0.11631317109642\\
0.0785548416905773	0.121468298926158\\
0.0746891678747084	0.125081846220933\\
0.070355719151612	0.129347943687506\\
0.0668937551397346	0.133613508276351\\
0.0644643864505321	0.136600588925574\\
0.0629318235248991	0.138263872008343\\
0.0621243300318304	0.139403618736192\\
0.061886765641771	0.1407824998113\\
};

\addplot [color=mycolor3, line width=2.0pt, mark=o, mark options={solid, mycolor3}]
  table[row sep=crcr]{%
0.0828147138598443	0.106838396170415\\
0.0822427210953426	0.106910750893781\\
0.0816918934287497	0.106988688044534\\
0.0811579746930352	0.107070170129455\\
0.080646177087105	0.107159251399691\\
0.080147532885883	0.107250469025299\\
0.0796673869334692	0.107348085977374\\
0.0792023901198964	0.10745033035391\\
0.0787530493717945	0.107558125520709\\
0.078315721330954	0.107669202847832\\
0.0778938771800309	0.107786782388739\\
0.0742628069112503	0.109201609764355\\
0.0717207831314339	0.110827253560147\\
0.0696303587461848	0.112805756431128\\
0.0678766122658472	0.115156231595961\\
0.0663880821097755	0.117916147488229\\
0.0651189335263523	0.121142539225142\\
0.0640403252545489	0.12491323017372\\
0.0631363664777498	0.129330722539092\\
0.0624115554689777	0.13455325946774\\
0.0618867744228747	0.140782516495479\\
};

\addplot [color=mycolor4, line width=2.0pt, mark=o, mark options={solid, mycolor4}]
  table[row sep=crcr]{%
0.151409576670851	0.12136040303444\\
0.136090255802373	0.121420509575487\\
0.131657083063191	0.121492028936449\\
0.129063023812588	0.12156261083609\\
0.127365712689238	0.12162707767165\\
0.126058671229292	0.121690953647524\\
0.125009232978169	0.121753926753975\\
0.124180270510127	0.121812784661968\\
0.123463470378846	0.121871753916279\\
0.122842975116097	0.121929912242517\\
0.122283151205337	0.121989290534354\\
0.118841831944344	0.122564976082608\\
0.116979973033949	0.123136946498038\\
0.115590103343297	0.123788075209681\\
0.114446223744868	0.12456094828032\\
0.113446498294809	0.125515177753491\\
0.112518449557615	0.12676856030202\\
0.111596367758191	0.128579842268232\\
0.110648420108789	0.131473039781555\\
0.10969996996944	0.13674794885341\\
0.109069318365017	0.149740199136381\\
};

\addplot [color=mycolor5, line width=2.0pt, mark=o, mark options={solid, mycolor5}]
  table[row sep=crcr]{%
0.151193662042105	0.121360133325397\\
0.150999398460756	0.121341648471462\\
0.150785375361693	0.121331579555364\\
0.150551859230439	0.121330490519544\\
0.150299214242748	0.121338959929293\\
0.150027899125396	0.121357578886534\\
0.14973846371635	0.121386948571094\\
0.149431543132863	0.121427677343149\\
0.149107850283421	0.121480377333707\\
0.148768166965275	0.121545660500008\\
0.148413326126204	0.121624133935766\\
0.144227880743575	0.123251695106887\\
0.139858071864182	0.126090201559387\\
0.134584280785424	0.129625149568039\\
0.127520304219712	0.13228319484995\\
0.119461181472578	0.133466481534249\\
0.113186147657725	0.134624791876236\\
0.109705454748647	0.137119525559907\\
0.108383823922112	0.140963257674786\\
0.108377813226434	0.145416971694456\\
0.109069344830635	0.149720028968287\\
};

\addplot [color=mycolor6, line width=2.0pt, mark=o, mark options={solid, mycolor6}]
  table[row sep=crcr]{%
0.151193385792455	0.121360133740345\\
0.149407106833375	0.121365291375979\\
0.147732371230288	0.121377779199997\\
0.146133018532024	0.121394647718879\\
0.144633358531863	0.121419262563249\\
0.143208597702057	0.121449254641837\\
0.141858944075378	0.121485495613517\\
0.140596374202924	0.121529678136428\\
0.139380004396293	0.121577822258444\\
0.138223096114159	0.121631714752656\\
0.137120815015898	0.121691232781498\\
0.12839952220889	0.12257376209711\\
0.123021306900885	0.123760677319839\\
0.119051337640885	0.125311483009311\\
0.11604332377842	0.127228680317272\\
0.113748926884188	0.129537851427539\\
0.111993200754682	0.132279988623032\\
0.110706224635557	0.135533467167716\\
0.109804978212715	0.139400415523482\\
0.109265292425034	0.144045788245942\\
0.109069336997397	0.149720020525056\\
};

\addplot [color=mycolor7, line width=2.0pt, mark=o, mark options={solid, mycolor7}]
  table[row sep=crcr]{%
0.19023930079081	0.185730375687659\\
};

\end{axis}
\end{tikzpicture}%
    \vspace{-1.cm}
  \centerline{(b) $K=3$} \medskip
\end{minipage}
\vspace{-0.4cm}
\caption{
Tradeoff (in terms of square root of \ac{crb}) between \ac{bp} and \ac{ms}: (a) $K=2$ (Targets 1 and 2); (b) $K=3$ (Targets 1, 2, and 3).}
\label{tradeoff}
\end{figure}

\subsubsection{Tradeoff between \ac{bp} and \ac{ms}}
In Fig. \ref{tradeoff}, the performance tradeoff between \ac{bp} and \ac{ms} is evaluated and compared across different schemes. Specifically, the results are examined for $K=2$ and $K=3$ targets. We observe that, except for the non-adaptive \ac{apa} scheme, all curves exhibit the fundamental tradeoff between \ac{bp} and \ac{ms}. Additionally, compared to \ac{cpa}, the \ac{fdb} schemes deliver significantly better performance in terms of the bistatic-monostatic performance tradeoff, owing to a higher degree of freedom in optimization. Furthermore, compared to the weighted-sum mismatch approaches (\ac{fdb}-WBF, \ac{fdb}-WVM, \ac{cpa}-WBF, and \ac{cpa}-WVM), the weighted-sum \ac{crb} approaches (\ac{fdb}-WCRB and \ac{cpa}-WCRB) achieve the best bistatic-monostatic performance tradeoff, as they yield weak Pareto boundaries. It is also worth noting that, compared to schemes relying on WBF, those relying on WVM consistently achieve a better bistatic-monostatic performance tradeoff, with performance very close to the weak Pareto boundary. Moreover, the performance gap between WBF and WVM becomes more pronounced in scenarios with more targets. The underlying insight is that approximating the variance matrix is more direct and effective in preserving the desired spatial characteristics of the transmit signal compared to approximating the beamformers, as the elements in the \ac{fim} are directly determined by the variance matrix.

\section{Conclusion}
In this paper, we characterized the tradeoff between \ac{bp} and \ac{ms} in a \ac{mimo}-\ac{ofdm} system. We derived the \acp{crb} for both paradigms and formulated a \ac{moo} problem. A weighted-sum \ac{crb} approach was proposed, ensuring the weak Pareto boundary. Additionally, we introduced and solved two mismatch-minimizing criteria based on beamformer mismatch and variance matrix mismatch. Numerical results demonstrated the performance tradeoff between the two paradigms and highlighted the superiority of the weighted-sum variance matrix mismatch approach, emphasizing the importance of balancing \ac{bp} and \ac{ms} in future \ac{isac} systems.

\appendices

\bibliographystyle{IEEEtran}
\bibliography{IEEEabrv,mybib}

\begin{thebibliography}{10}
\providecommand{\url}[1]{#1}
\csname url@samestyle\endcsname
\providecommand{\newblock}{\relax}
\providecommand{\bibinfo}[2]{#2}
\providecommand{\BIBentrySTDinterwordspacing}{\spaceskip=0pt\relax}
\providecommand{\BIBentryALTinterwordstretchfactor}{4}
\providecommand{\BIBentryALTinterwordspacing}{\spaceskip=\fontdimen2\font plus
\BIBentryALTinterwordstretchfactor\fontdimen3\font minus \fontdimen4\font\relax}
\providecommand{\BIBforeignlanguage}[2]{{%
\expandafter\ifx\csname l@#1\endcsname\relax
\typeout{** WARNING: IEEEtran.bst: No hyphenation pattern has been}%
\typeout{** loaded for the language `#1'. Using the pattern for}%
\typeout{** the default language instead.}%
\else
\language=\csname l@#1\endcsname
\fi
#2}}
\providecommand{\BIBdecl}{\relax}
\BIBdecl

\bibitem{fan2022jsac}
F.~Liu \emph{et~al.}, ``Integrated sensing and communications: Toward dual-functional wireless networks for 6{G} and beyond,'' \emph{IEEE Journal on Selected Areas in Communications}, vol.~40, no.~6, pp. 1728--1767, 2022.

\bibitem{hui2022st}
H.~Chen \emph{et~al.}, ``A tutorial on terahertz-band localization for 6{G} communication systems,'' \emph{IEEE Communications Surveys and Tutorials}, vol.~24, no.~3, pp. 1780--1815, 2022.

\bibitem{henk2019twc}
R.~Mendrzik \emph{et~al.}, ``Harnessing {NLOS} components for position and orientation estimation in {5G} millimeter wave mimo,'' \emph{IEEE Transactions on Wireless Communications}, vol.~18, no.~1, pp. 93--107, 2019.

\bibitem{furkan2022tvt}
M.~F. Keskin \emph{et~al.}, ``Optimal spatial signal design for mmwave positioning under imperfect synchronization,'' \emph{IEEE Transactions on Vehicular Technology}, vol.~71, no.~5, pp. 5558--5563, 2022.

\bibitem{fan2018tsp}
F.~Liu \emph{et~al.}, ``Toward dual-functional radar-communication systems: Optimal waveform design,'' \emph{IEEE Trans. Signal Process.}, vol.~66, no.~16, pp. 4264--4279, 2018.

\bibitem{fan2018twc}
------, ``{MU-MIMO} communications with {MIMO} radar: From co-existence to joint transmission,'' \emph{IEEE Trans. Wireless Commun.}, vol.~17, no.~4, pp. 2755--2770, 2018.

\bibitem{fan2022tsp}
------, ``{Cram\'{e}r-Rao} bound optimization for joint radar-communication beamforming,'' \emph{IEEE Trans. Signal Process.}, vol.~70, pp. 240--253, 2022.

\bibitem{hua2023optimal}
H.~Hua \emph{et~al.}, ``Optimal transmit beamforming for integrated sensing and communication,'' \emph{IEEE Transactions on Vehicular Technology}, vol.~72, no.~8, pp. 10\,588--10\,603, 2023.

\bibitem{he2023full}
Z.~He \emph{et~al.}, ``Full-duplex communication for {ISAC}: {Joint} beamforming and power optimization,'' \emph{IEEE Journal on Selected Areas in Communications}, vol.~41, no.~9, pp. 2920--2936, 2023.

\bibitem{rivetti2023spatial}
S.~Rivetti \emph{et~al.}, ``Spatial signal design for positioning via end-to-end learning,'' \emph{IEEE Wireless Communications Letters}, vol.~12, no.~3, pp. 525--529, 2023.

\bibitem{yu2023globecom}
Y.~Ge \emph{et~al.}, ``Integrated monostatic and bistatic mm{W}ave sensing,'' in \emph{GLOBECOM 2023 - 2023 IEEE Global Communications Conference}, 2023, pp. 3897--3903.

\bibitem{henk2022jstsp}
A.~Fascista \emph{et~al.}, ``{RIS}-aided joint localization and synchronization with a single-antenna receiver: Beamforming design and low-complexity estimation,'' \emph{IEEE Journal of Selected Topics in Signal Processing}, vol.~16, no.~5, pp. 1141--1156, 2022.

\bibitem{ehrgott2005multicriteria}
M.~Ehrgott, \emph{Multicriteria optimization}.\hskip 1em plus 0.5em minus 0.4em\relax Springer Science \& Business Media, 2005, vol. 491.

\bibitem{tom2010spm}
Z.-Q. Luo \emph{et~al.}, ``Semidefinite relaxation of quadratic optimization problems,'' \emph{IEEE Signal Processing Magazine}, vol.~27, no.~3, pp. 20--34, 2010.

\bibitem{youli2024twc}
L.~You \emph{et~al.}, ``Integrated communications and localization for massive {MIMO LEO} satellite systems,'' \emph{IEEE Transactions on Wireless Communications}, vol.~23, no.~9, pp. 11\,061--11\,075, 2024.

\bibitem{fortin2004trust}
C.~Fortin \emph{et~al.}, ``The trust region subproblem and semidefinite programming,'' \emph{Optimization methods and software}, vol.~19, no.~1, pp. 41--67, 2004.

\end{thebibliography}

\end{document}